\documentclass{revtex4-1}
\usepackage{graphicx}
\usepackage{latexsym}
\usepackage{amsmath}
\usepackage{bm}
\usepackage{amssymb}

\newcommand{\be}{\begin{equation}}
\newcommand{\ee}{\end{equation}}
\newcommand{\bea}{\begin{eqnarray}}
\newcommand{\eea}{\end{eqnarray}}
\begin{document}
\title{Induced polar perturbations  in relativistic stars with stochastic 
effects in the dense matter at sub-hydro mesoscopic scales: A theoretical
probe at intermediate length scales }
\author{Seema Satin}
\affiliation{Indian Institute for Science Education and Research, Kolkata, India}
\email{seemasatin@iiserkol.ac.in}

\begin{abstract}
A linear response  relation between metric and fluid perturbations driven by
a background noise source is used as a framework for obtaining non-radial
polar perturbations in dense matter relativistic stars. The perturbations
carry a generalized stochastic nature as solutions to the classical
Einstein-Langevin equation which has been recently proposed. The
significance of these stochastic non-radial polar perturbations lies at 
probing the intermediate sub-hydro scales inside the dense fluid. This 
study extends towards a  non-equilibrium/near-equilibrium 
statistical mechanics study for relativistic star interiors. 
We address the non-radial polar perturbations in stars which are 
important from the point of view of detection in future.  
A generalized stochastic noise  which originates as the remanant of collpase
mechanism in isolated star is expected to give  rise to 
such stochastic polar perturbations at intermediate sub-hydro scales. 
 More specifically it is the interplay between the
degeneracy pressure of exotic matter and  the gravitational pressure that
 gives rise to
the seeds of stochastic effects or noise in the background of the gravitating
body towards the  near-equilibrium configuration.
\end{abstract}
\maketitle
\section{Introduction}
Spherical polar perturbations in relativistic stars are of significance for the
analysis of pulsation modes of the fluid matter \cite{nilskok,nil}. The current 
efforts in this direction are towards simulations and detection 
\cite{marck,abbot,radice} through gravitational waves, the
modes of oscillations in systems like the coalescence of two neutron stars or a 
core collapse in a supernovae. 
One of the aims to study perturbations of relativistic stars is to 
know the equation of state that the interiors of the compact objects
are composed of. This is an open area, with nuclear physics and astroparticle
details  to model the quantum matter inside these stars \cite{andreas}. How
 does this nuclear matter inside the compact object under the
 influence of gravity behave, is thus an open question which also opens up 
possibilities  for
 new theoretical constructs. Often one models the matter as  quantum fluid, 
the details of which are being currently investigated at  either the 
microscopic scales or the macroscopic hydrodynamic scales \cite{shternin}. 
Transport properties from nuclear physics details and with microscopic
details have been known to be of interest in this regard. However  
accurate formalism to model complexity of the extreme conditions
 inside the compact stars is still desirable. The scales that are being 
investigated at present, namely the micro and macro scales are different than
 we wish to explore.
We wish to address phenomena at sub-hydo mesoscopic scales  such that
 strong gravity effects on the 
nuclear matter in bulk may be related to transport properties at these 
intermediate scales using statistical analysis. Thus our formalism adds onto 
the scales presently under focus in this area. 

With this aim, recently a new theoretical framework based on the classical
 Einstein-Langevin equation has been under  consideration and formulation
 \cite{seema1,seema2,seema3, seema4}. 
This may also help to improve our understanding of how the cross over from 
quantum microscopic scales to that at
 macroscopic scales occurs and filters out essential interesting 
characteristics. Another interesting and important aspect will be that of
understanding the scales at which  quantum fluid inside compact objects
begins to get
influenced by strong gravity in its bulk variables and the specific 
characteristics that it shows up. We will thus model for large fluid particle
size in our formulations as we have described further in the article to be
considered for probing such properties.  
The question  about
 scales at which  strong gravity starts showing its effects on quantum matter
 in bulk, is hence interesting to explore.
These intermediate scales are expected to lie on the verge  where strong gravity
starts showing its effect on the dense matter at least perturbatively, in a 
relativistic star. 
 The characteristics of fluid at these scales in cold dense matter can be
 known from the stochastic mode analysis. With 
the new approach based on a classical Einstein-Langevin equation 
 one can touch upon these very scales.   

In this article we work out the basic expressions for stochastic polar 
perturbations induced by noise or internal sources in the dense fluid 
 for a non-rotating spherically symmetric star. 
The radial perturbations in this context have been worked out in \cite{seema3,
seema4}. 
 This article marks the beginning of
 the study towards non-radial stochastic perturbations, presenting basic
expressions in analytical form for which numerical solutions
 will follow in future. Our theoretical formulations add to the
 usual pulsation modes studied in asteroseismology, these
 new stochastic effects which encode the more refined mesoscopic physics
of the relativistic fluids.   
It is the degeneracy pressure and its generalized random \cite{seema3,seema4}
 fluctuations that are at play as noise in the background,
whose cumulative effect on the polar perturbations  are expected to 
affect  the pulsation modes of the fluid at intermediate scales. 
This direction in research has the potential to add on a new framework  to
 the ongoing efforts in numerical relativity \cite{nils3,Nilsallen} with 
an overlap of similar theoretical constructs for relativistic fluids. 
  As we will see towards the end of the article, the complexity of the 
analytical expressions calls of devising specific
numerical methods to evaluate  and solve them  for
use in asteroseismology, it gives quite a large scope
to numerical relativists to devise techniques here. With this efforts
on probing more refined classical picture in terms of equilibrium
and non-equilibrium statistical properties of the dense fluid,  a 
mesoscopic range theory is underway.  As a first step which
is the simplest in this regard, we have considered
non-rotating spherically symmetric configuration. The basic first
principles  devised here  are an advancement in the theoretical formulations
and form  the first set of new results for non-radial perturbations.  

We begin with reviewing the classical Einstein-Langevin equation,
\section{The Einstein-Langevin equation}
 A linear response relation for perturbations of relativistic stars has been
introduced in \cite{seema3}, which takes the form of a classical
 Einstein-Langevin equation.
\be \label{eq:ltreq}
 \delta G_{\alpha \beta}[h;x) - 8 \pi \delta T_{\alpha \beta}[h;x) - 8 \pi
 \int \mathcal{R}(x-x') \delta T_{\alpha \beta}[\xi]; x')
= \tau_{\alpha \beta}[g;x)
\ee
where $h$ denotes the metric perturbations and $\xi$, the fluid perturbations.
The term $\tau_{\alpha \beta}[g;x) $ denotes the stochastic source inside
 the dense matter of the relativistic star.
This is defined as $\tau_{\alpha \beta}(x) = \delta_s T_{\alpha \beta}(x) $,
 and denotes the  generalized stochastic fluctuations of  matter. 
 Thus the Langevin noise is defined with $\langle \tau_{\alpha \beta}(x)
 \rangle  = 0 $ and the two point correlation
 $ \langle \tau_{\alpha \beta}(x) \tau_{\delta \gamma}(x') \rangle $ 
 meaningfully. Note that $\tau_{\alpha \beta} $ is defined on
 the background unperturbed metric $g_{\alpha \beta}(x) $.  
The linearized perturbed Einstein's equations are  covaraintly  conserved
 w.r.t the background metric $g_{\alpha \beta}$, while the noise
is also covariantly conserved w.r.t the background metric, hence
 $ \nabla_\alpha \tau^{\alpha \beta}(x) = 0 $. 

In the above equation $\mathcal{R}(x-x')$ is the response kernel which
 connects the metric and fluid perturbations. As prescribed in \cite{seema3},
 this is delta correlated  in the spacetime variables  
$\mathcal{R}(x-x') = \delta(x-x')$ for the  case of a 
vanishing noise component in the set of equations, or else for external 
sources which perturb the system in a deterministic fashion where 
$\tau_{\alpha \beta} $.
One can question upon the fluid sources if they are strong enough to perturb 
the metric. Here we propose to model for fluid particle size as an element of 
the order of few 10s of meters to  100 meters  inside the dense star.
The  generalized fluctuations of the degenerate pressure are then large towards
the end state of collapse is expected in our modelling, and it is the 
cumulative effect of these background fluctuations that is effective enough
to perturb or shift trajectories of the fluid. 
 
We assign
such  stochastic sources to arise due to the remanant effects of 
collapse phenomena and complex mechanical effects in the interiors of dense 
matter for the fluid 
particle size of the order that we mention later and the sub-hydro scales
which for the large particle size lie a little below the macroscopic hydro 
scales. Thus it
is not the thermal phenomena that we model for, but mechanical effects
in cold matter due to various other possibilities  like the
escape of neutrinos from the bulk, the pressure effects due to layers 
pressing each other in the collapse etc. Even for non-rotating 
spherically symmetric star collapse, we can assume small non-radial 
fluctuations as the layers of degenerate matter press onto each other 
at various depths under the effect of  strong gravity , though we do not 
model frictional sources here.  
This is expected  to have significant impact which are large enough then
to perturb  the system stochastically. Our sub-hydro scales
are  then of the order of this estimate.  Given the compact 
matter and its dense nature it is these scales that are yet unexplored, which
we intend to touch upon. Thus one can veiw  such 
large fluid particles as  showing generalized mechanical fluctuations 
(which may even be due to quantum mechanical fluctuations persisting in the 
bulk of quantum matter) and an ensemble of
these for a stochastic analysis, such that the impact on the system 
can be seen as mesoscopic perturbations. 
\subsection{The model of relativistic star}
The equilibrium configuration of a relativistic spherically symmetric star
is described by a coordinate system $(t,r,\theta,\phi)$, w.r.t which the 
geometry of spacetime takes the form
\be
ds^2 = - e^{2 \nu} dt^2 + e^{2 \lambda} dr^2 + r^2 (d \theta^2 + \sin^2 \theta
 d \phi^2)
\ee 
 where $\lambda$ is related to the mass inside the radius $r$ , by
$ e^{2 \lambda} = ( 1-2 m/r)^{-1} $. 
The matter is modelled by the perfect fluid stress tensor
\be
T_{\alpha \beta} (x)= u_\alpha u_\beta (\epsilon+p) + g_{\alpha \beta} p
\ee
The TOV equation of hydrostatic equilibrium is given by
\be
\frac{dp}{dr} = -  \frac{(\epsilon+ p) (m + 4 \pi r^3 p)}{r (r- 2 m)}  
\ee 
the adiabatic index which governs the response of the stellar material to 
pulsational compressions is given by
\be
\Gamma_1 = [(\epsilon+ p)/p] \frac{d p}{d \epsilon} .
\ee
 
\subsection{Model of noise for polar perturbations}
We have  defined the source  $\tau_{\alpha \beta}(x)$ in the E-L equation as
a stochastic term with an explicit form $\tau_{\alpha \beta} =\delta_s
T_{\alpha \beta} $ , where 's' in $\delta_s $ denotes stochastic. We precribe a
model in which the generalized stochasticity arises  in the pressure variable
such that $\delta_s p(r,t) = \delta_s p(r) e^{i \omega t} $ acts as the  
 seeds of perturbations cumulatively. This happens towards the end
stages of collapse when a collapsing star reaches
 a near-static configuration. Hence our background which is spherically
 symmetric is the equilibrium, around which we seek solutions.    
For inducing polar perturbations the polar decomposition of noise 
 $\tau_{\alpha\beta} $ as a pull back on the on the 2-sphere 
 takes the form  
\be
\tau_{ab} = e_{ab} \tau_{lm}^{scalar} Y_{lm}+  \tau_{lm}^{polar}
 \nabla_a \nabla_b Y_{lm}
\ee
where $e_{ab}$ is the metric restricted to the 2-sphere. 
The indiced $a,b $ are reserved for $\theta, \phi $, while $\alpha, \beta $ run 
 for $\{t,r,\theta, \phi \} $ . 
In the Regge-Wheeler gauge $\tau_{lm}^{polar} =0 $ corresponding to the
decomposition of the metric (see appendix).
The projection for $t-t$ and $r-r$ components on the 2-sphere can be given by,
\bea
& & \tau^t_t Y_{lm} = \sum_{lm} \delta_s \epsilon(r)e^{i \omega t} Y_{lm}(\theta,\phi)  ; 
\tau^r_r Y_{lm} = \sum_{lm}
 \delta_s p (r) e^{i \omega t} Y_{lm}(\theta,\phi) 
\eea
where $\delta_s \epsilon (r,t), \delta_s p(r,t) $ are the 
generalized random fluctuations in pressure and energy density.  
The $\theta,\phi$ components have the form, 
\be
 \tau^\theta_\theta(r,t) Y_{lm}(\theta,\phi) =    
 \tau^\phi_\phi(r,t) Y_{lm}(\theta,\phi) = \sum_{lm}  \delta_s p_{lm}(r,t)
 Y_{lm}(\theta,\phi)
\ee 
As we see further, these form the sources of induced perturbations .   
Evaluating in this article for $m=0$ case, the relevant form of
 expressions accordingly can be written as
\bea
& & \tau^t_t =  \delta_s \epsilon(r)e^{i \omega t} P_l(\cos \theta)  ; 
\tau^r_r =  \delta_s p(r) e^{i \omega t} P_l(\cos \theta) 
\label{eq:tt1}\\
& & \tau^\theta_\theta = \tau^\phi_\phi = \sum_l \delta_s p_l(r) 
e^{i \omega t} P_l(\cos \theta).
\eea 
where we have assumed the seeds of fluctuations to take the stochastic
harmonic dependence $e^{i \omega t} $ , where $\omega $ has 
 a random nature with a distribution $P(\omega)$, occurring due to
 uncertainty in determining its exact value. 
Note the $l$ dependence of the random pressure fluctuations on the sphere for 
the $\theta,\phi$ components at a radial depth $r$ in the star. This is the
 basic new construct of our noise model
to account for polar randomness on the 2-sphere. While the projection for the
$t,r$ components on the sphere is given by equation (\ref{eq:tt1}).  

The induced perturbations due to such a noise as source, can be obtained from
 the
solutions of the classical Einstein-Langevin equation as we show in the 
subsection below. 
\section{Induced stochastic polar perturbations with the Einstein-Langevin
formalism}   
A review of polar perturbations in a spherically symmetric star is given in
the appendix. 

Polar perturbations of relativistic stars are obtained in literature often
using the Regge Wheeler gauge where the perturbed metric functions 
 $H_0(r,t), H_1(r,t) $ and $K(r,t)$ remain independent and non-zero.  
In our formalism these perturbations attain a generalized stochastic nature,
 and remain functions of $r,t$ but with a different dependence
 $e^{\gamma t}$ where $\gamma$ is of complex random nature. The 
randomness comes from uncertainty in the exact value of the frequency.The
 nature of this complex frequency is such that $\gamma = 
\gamma_R - i\gamma_I$ where $\gamma_R $ and $\gamma_I$ are
the real and imaginary parts, each of  which are random in nature due to
lack of determining the exact value. Thus they have a probability
distritution  which may be a joint probablitiy distribution for the
real and imaginary parts $P(\gamma_R, \gamma_I) $.   
We proceed to obtain solutions for the Einstein-Langevin equation
in terms of these potentials.
Assuming  the Regge Wheeler gauge,
\bea
0 &= & \delta G^\theta_\theta - \delta G^\phi_\phi =(H_2 - H_0)
\frac{1}{2 r^2}(\partial_\theta^2 - \cot \theta \partial_\theta - 
\frac{1}{\sin^2 \theta}\partial^2_\phi ) Y_{lm}
\eea
or $ H_0 = H_2 $. 
From 
\be
\delta G_{r \theta} = 0
\ee 
one gets
\be
e^{-2 \nu} \partial_t H_1 = H'_0 + 2 \nu' H_0 - K'
\ee
The non-zero components of the Einstein-Langevin equations then take the
following form, with $\mathcal{R}(x-x') = {\mathcal{R}_1}(t-t')\delta(r-r')
\delta(\theta- \theta') \delta(\phi - \phi')$ and $\tilde{\mathcal{R}}
(\gamma) = \int \mathcal{R}_1(\tau) e^{- i \gamma \tau} d \tau $ defined as 
Laplace transform which gives susceptibility for the system to get perturbed. 

The $t-t$ component of the E-L (Einstein-Langevin) equation is given by,
\be
\delta G^t_t [h,x) -8 \pi  \delta T^t_t[h,x) - 8 \pi \int \mathcal{R}(x-x') 
\delta T^t_t [\xi,x') d^4x' = \tau^t_t[g,x)  
\ee
which with the polar decomposition reads, 
\begin{widetext}
\bea \label{eq:tt}
 & & [ \{- e^{-2 \lambda} \partial_r^2 - e^{-2 \lambda} ( \frac{3}{r} -
 \lambda') \partial_r + [ \frac{1}{2 r^2} (l-1)(l+2) + 8 \pi (\epsilon+p)] 
 \} K +\{ \frac{e^{-2 \lambda}}{r} \partial_r - e^{-2 \lambda} \nonumber \\
& & ( \frac{1}{r^2} - \frac{2}{r} \lambda ' + \frac{e^{2 \lambda}}{ 2 r^2}
 l(l+1)) + 4 \pi (\epsilon + p) \} H_0 +  \tilde{\mathcal{R}}(\gamma) 8 \pi 
(\frac{e^{-\lambda}}{r^2} ( (\epsilon+p)\partial_r + \epsilon') W 
+  (\epsilon+p) \frac{l(l+1)}{r^2} V) ] P_l (\cos \theta)= \nonumber  \\
& & -\delta_s
\epsilon(r) e^{i \omega t} P_{\tilde{l}}(\cos \theta)
\eea
The $t-r$ component which has no source term reads,
\be
\delta G^r_t [h,x) -8 \pi  \delta T^r_t[h;\xi, x)   = 0 
\ee
given by
\be \label{eq:tr}
[e^{- \lambda + \nu} (\partial_r + 2 \nu') H_0 - e^{-\lambda + \nu}
 \partial_r K - (\epsilon+ p) \frac{e^{-2 \nu + \lambda}}{r^2} \partial_t W]
P_l(\cos \theta)
=0 
\ee
The $r-r$ component is given by,
\be
\delta G^r_r [h,x) -8 \pi  \delta T^r_r[h,x) - 8 \pi \int \mathcal{R}(x-x')
 \delta T^r_r [\xi,x') d^4x' = \tau^r_r[g,x)  
\ee
with its projection on the two sphere reads,
\bea \label{eq:rr}
& & [ \{e^{-2 \nu} \partial_t^2 + e^{-2 \lambda} (\frac{1}{r} - \nu') 
\partial_r - 8 \pi \Gamma_1 p + \frac{1}{r} (l-1)(l-2) \} K  - 
\{ \frac{e^{-2 \lambda}}{r} + \frac{1}{r^2} ( e^{-2 \lambda} -1) 
 + \frac{1}{r} (l-1)(l+2) +  \nonumber\\
& & 4 \pi \Gamma_1 p \} H_0  
 - 8 \pi \tilde{\mathcal{R}}(\gamma) [ \Gamma_1 p \frac{e^{-\lambda}}{r^2}
 \partial_r + \frac{p'}{r^2} e^{-\lambda}] W - 
 8 \pi \tilde{\mathcal{R}}(\gamma) \Gamma_1 p \frac{l(l+1)}{r^2} V ] 
P_l (\cos \theta) = \delta_s p(r) e^{i \omega t} P_{\tilde{l}}
 (\cos \theta).
\eea

The $\theta-\theta $ component is given by,
\be 
\delta G^\theta_\theta [h,x) -8 \pi  \delta T^\theta_\theta[h,x) - 8 \pi 
\int \mathcal{R}(x-x') \delta T^\theta_\theta [\xi,x') d^4x' = \tau^\theta_\theta[g,x)  
\ee
which reads,
\bea \label{eq:thetatheta}
& &[ [e^{-2\nu} \partial_t^2 - \frac{e^{-2 \lambda}}{2} \partial_r^2 -
 \frac{1}{2}
e^{-2 \lambda} ( \frac{2}{r} \nu' - 2\lambda' + \frac{1}{r}- 2 e^{2 \lambda}
\partial_r) \partial_r - 8 \pi \Gamma_1 p ] K \nonumber \\
& & [ - e^{-2 \lambda} ( \partial_r - \frac{1}{r} - \lambda') ( \partial_r + 
2 \nu') + \frac{1}{2} e^{-2 \nu} \partial_t^2 + \frac{1}{2} e^{-2 \lambda} 
\partial_r^2 + \frac{e^{-2 \lambda}}{2} ( \frac{2}{r} + \nonumber \\
& &  3 \nu' - \lambda') \partial_r - 8 \pi p - 4 \pi \Gamma_1 p ] H_0
+ 8 \pi \mathcal{R}(\gamma) [ \Gamma_1 p \frac{e^{- \lambda}}{r^2} \partial_r 
+ \frac{p'}{r^2} e^{-\lambda} ] W +  \nonumber \\
& & 8 \pi \mathcal{R}(\gamma) \Gamma_1 p \frac{l(l+1)}{r^2} V] P_l 
(\cos \theta) = 
 \delta_s p_{\tilde{l}}(r) e^{i \omega t} P_{\tilde{l}}(\cos \theta)
\eea
\end{widetext}
Thus we have four unknowns, $H_0,K,W,V$ and the four equations (\ref{eq:tt}),
(\ref{eq:tr}), (\ref{eq:rr}) and (\ref{eq:thetatheta}) which need  to be solved
numerically. One may also separate the
metric perturbations $H_0, K$ and the fluid matter  perturbations $W,V$
and write the fluid perturbations in terms of those of the metric.
 Note that $l$ on the lhs and $\tilde{l}$ on the
rhs of the above equations are different sets. While $\tilde{l} $ stand for the
 unperturbed background $g_{\alpha \beta}$,  $l$ stands for the  perturbed
 configuration on $ g_{\alpha \beta} + h_{\alpha \beta} $.
It is simple to see that $ \langle H_0 \rangle= \langle K \rangle= 
\langle W \rangle = \langle V \rangle = 0 $, as expected due
to the Langevin property of the noise which induces them, 
while it is the two point correlations which specify their forms including
the rms values to give estimates of the magnitudes  for these
perturbations. 
 
We can also go a step further analytically and
 from equation (\ref{eq:tr})  one can easily  obtain
\be
W =  \frac{1}{\gamma} \frac{r^2}{(\epsilon + p)} e^{3 \nu - 2 \lambda}
[(\partial_r + 2 \nu')H_0 - \partial_r K]
\ee  
(note the explicit appearance of $\gamma$ in the expressions from here ).
Using this and equation (\ref{eq:tt}) one can obtain the expression for $V$
after a simple exercise and substituing for $W$ and $V$ in equations
(\ref{eq:rr}) gives,  
\begin{widetext}
\bea \label{eq:rr2}
& &[ \{ \{ [- \frac{e^{-2 \lambda}}{r}(\partial_r + 2 \nu') - 
\frac{(l-1)(l+2)}{2 r^2} + \frac{ e^{-2 \lambda -1} }{r^2} - 4 \pi \Gamma_1 p
+ 8 \pi \mathcal{R}(\gamma) \frac{e^{3(\nu -\lambda)}}{i\omega} [ \Gamma_1 p \{
\frac{2}{r} + 3 \nu' - 2 \lambda' - \frac{\epsilon'}{(\epsilon+p)} \}\nonumber 
\\
& & ( \partial_r + 2 \nu') + ( \partial_r^2 + 2 \nu'') + 2 \nu' \partial_r ]
-\frac{\Gamma_1 p}{(\epsilon+p)} \frac{e^{-2 \lambda}}{r} \partial_r -
 e^{- 2 \lambda}( \frac{1}{r^2} - \frac{2}{r} \lambda' + 
\frac{ e^{2 \lambda}}{2 r^2} l (l+1) + 4 \pi (\epsilon+p) )] + 
 \nonumber \\
 & & 8 \pi e^{3(\nu- \lambda)} \frac{\mathcal{R}(\gamma)}{ \gamma}
 \{(\frac{2}{r}  
 + 4 \nu' - 2 \lambda') (\partial_r + 2 \nu') - \partial_r^2 + 2 \nu'' + 
2 \nu' \partial_r \} \} H_0
+ [ \{ e^{-2\nu} \partial_t^2 + e^{-2 \lambda} ( \frac{1}{r} - \nu') \partial_r
+ \frac{1}{2 r^2} (l-1)(l+2) - \nonumber\\
& &  8 \pi \Gamma_1 p - 8 \pi \tilde{\mathcal{R}} (\gamma)
\frac{ \Gamma_1 p}{\gamma ( \epsilon+ p)} e^{3(\nu- \lambda)} [
\{\frac{2}{r} + 4 \nu' - 2 \lambda' - \frac{\epsilon'}{(\epsilon+p)} \}
\partial_r  - \partial_r^2 ] + [e^{-2 \lambda} \partial_r^2 + 
 ( \frac{3}{r} -\lambda') e^{- 2 \lambda} \partial_r - \nonumber \\
& &  \frac{1}{2} (l-1)(l+2)- 8 \pi
(\epsilon+ p)+ 8 \pi e^{3(\nu-\lambda)} \frac{\mathcal{R}(\gamma)}{\gamma} 
\{ 2 \frac{\partial_r }{r} + 4 (\nu' - 2 \lambda')\partial_r + 
  \partial_r^2 \}] K \}]] P_l(\cos\theta) = \nonumber \\
& &  (\delta_s p(r) - \frac{\Gamma_1 p}{(\epsilon+p)} \partial_s \epsilon(r) )
e^{i \omega t} P_{\tilde{l}} (\cos \theta).  
\eea
From equation (\ref{eq:thetatheta}),
\bea \label{eq:thth2}
& &[ \frac{1}{2} e^{-2 \nu} \partial_t^2 - \frac{1}{2} e^{- 2 \lambda}
 \partial_r^2 - \frac{1}{2} e^{- 2 \lambda} ( \frac{1}{r} + \nu' -
 \partial_r ) \partial_r - 8 \pi \Gamma p  - e^{3(\nu- \lambda)}
 \frac{8 \pi \tilde{\mathcal{R}}(\gamma) \Gamma_1 p }{ \gamma (\epsilon+p)}
 \nonumber \\
& &  [ \{ \frac{2}{r} + 4 \nu' - 2 \lambda' - \frac{\epsilon'}{(\epsilon+p)}
 \} \partial_r - \partial_r^2 ] + \frac{\Gamma_1 p}{(\epsilon+p)} 
[ - e^{- 2 \lambda} \partial_r^2 - (\frac{3}{r} - \lambda' )e^{- 2 \lambda} 
\partial_r + \nonumber \\
& & \frac{1}{2 r} (l-1)(l+2) + 8 \pi (\epsilon+p) - 8 \pi \mathcal{R}(\gamma)
 \frac{e^{3(\nu- \lambda)}}{ \gamma} \{ 2 \frac{\partial_r}{r} + 
 (4 \nu' - 2 \lambda') \partial_r + \partial_r^2\}]
K + \nonumber \\
& & \{ \{ \frac{1}{2} e^{- 2 \nu} \partial_t^2 - \frac{1}{2} e^{- 2 \lambda} 
\partial_r^2 + e^{ - 2 \lambda} ( \frac{1}{r} + \frac{\nu'}{2}  
 \frac{ \lambda'}{2}) \partial_r + 2 \nu' \lambda' e^{-2 \lambda} + 8 \pi (1-
\frac{\Gamma_1}{2} \} \nonumber \\
& &  - 8 \pi \tilde{\mathcal{R}}(\gamma) \frac{e^{3(\nu - \lambda)}}{\gamma}
[ \Gamma_1 p \{ \frac{2}{r} + 3 \nu' - 2 \lambda' - 
 \frac{\epsilon'}{( \epsilon+p)} \}(\partial_r + 2 \nu') + 
( \partial_r^2 + 2 \nu') + \nonumber\\
& &  (\partial_r^2 + 2 \nu'') + 2 \nu' \partial_r ] + 
\frac{\Gamma_1 p}{(\epsilon + p)} 
[e^{- 2 \lambda}{r} \partial_r - e^{-2 \lambda} ( \frac{1}{r^2} - \frac{2}{r}
\lambda' + \frac{e^{2 \lambda}}{2 r^2} l (l+1) + \nonumber \\
& & 4 \pi (\epsilon+p) + 8 \pi 
e^{3 ( \nu- \lambda) }\frac{\tilde{\mathcal{R}}(\gamma)}{\gamma} 
\{ \frac{2}{r} +
 4 \nu' - 2 \lambda')( \partial_r + 2 \nu') + \partial_r^2 + 2 \nu'' +
 2 \nu' \partial_r \} ] \} H_0] P_l \cos(\theta) = \nonumber \\
& & ( \delta_s p_{\tilde{l}}(r) + \frac{\Gamma_1 p}{(\epsilon+p)} 
\delta_s \epsilon(r)) e^{ i \omega t} P_{\tilde{l}} (\cos \theta)  
\eea 
\end{widetext}
Equations (\ref{eq:rr2}) and (\ref{eq:thth2}) can be written in a compact
form as
\be  \label{eq:x}
(X^1_l K + X^2_l H_0 ) P_l (\cos \theta)  = [\delta_s p(r)- \Gamma_1 
\frac{p}{(\epsilon + p)} \delta_s \epsilon(r)] e^{i \omega t}
  P_{\tilde{l}} (\cos (\theta))
\ee 
and 
\be \label{eq:y}
(Y^1_l K + Y^2_l H_0 ) P_l (\cos \theta) = [\delta_s p_{\tilde{l}}(r) + 
\Gamma_1 \frac{p(r)}{(\epsilon+p)} \delta_s \epsilon(r)]e^{i \omega t}
 P_{\tilde{l}}(\cos(\theta))
\ee
As the perturbations $K, H_0 $ are generalized random in nature also with
$\gamma$ being random  in our formulations, one is interested in two point 
correlations or rms values for these. We emphasise that point separated two
point correlations of the form $\langle K(r_1,t_1) K(r_2,t_2) \rangle,
\langle H(r_1,t_1) H(r_2,t_2) \rangle, \langle K(r_1,t_1) H(r_2,t_2) 
\rangle $ and $\langle H(r_1,t_1) K(r_2,t_2) \rangle $ once evaluated 
numerically will give insight to extended properties of dense matter with large
separations $r_2 - r_1 $ taking care of the causal effects inside the dense 
star.  Details on this aspect have been discussed in \cite{seema3,seema4} for
the radial perturbations where analytical form of final expressions
is easily possible. In the present case, we leave more detailed
describtion for a later stage when numerical  techniques for evaluation will
be investigated to progress further on the  accuracy of such details. In the 
present article we have presented  qualitatively the analysis of 
 expressions which design the theoretical base.  
 
The above equations need numerical solutions given the complex
analytical form,  to evaluate explicitly  for the functions $H_0$ and $K$.
As  one is interested here in the two point correlations, 
several expressions for two point correlations from above may also be 
obtained, one such example can be given by writing, 
\begin{widetext}
\bea
& & [X^1_{l_1} Y^1_{l_2} \langle K(r_1,t_1)K(r_2,t_2) \rangle + 
X^1_{l_1}Y^2_{l_2} \langle K(r_1,t_1) H_0(r_2,t_2) \rangle + X^2_{l_1}
 Y^1_{l_2} \langle H_0(r_1,t_1) K(r_1,t_2) \rangle  + \nonumber \\
& & X^2_{l_1} Y^2_{l_2} \langle H_0(r_1,t_1) K(r_2,t_2) \rangle] P_{l_1} 
( \cos \theta_1) P_{l_2}(\cos \theta_2)
 =  \langle \delta_s p(r_1) \delta_s p_{\tilde{l}}(r_2) \rangle
 e^{i \omega (t_1-t_2)} + \Gamma_1 p(r_2) 
\langle \delta_s p(r_1) \delta_s \epsilon(r_2) \rangle e^{i \omega(t_1
 - t_2)} - \nonumber \\
& & \Gamma_1 p(r_1)  \langle \delta_s \epsilon (r_1) \delta_s p_{l}(r_2)\rangle 
e^{i \omega (t_1- t_2)} - \Gamma_1^2 p(r_1) p(r_2) 
\langle \delta_s \epsilon(r_1) \delta_s \epsilon(r_2) \rangle 
 e^{i \omega(t_1- t_2)}
 \eea
\end{widetext}
With this, we have given the basic new expressions for stochastic polar 
perturbations for a spherically symmetric non-rotating configuration of a
star, which towards the end state of its collapse  carries
noise in terms of stochastic fluctuations of the stress energy tensor given by 
$\tau_{\alpha \beta}$.
\subsection{Cowling approximation}
In the Cowling approximation one neglects coupling of perturbations of metric
to the matter fields which is valid for short wavelengths. Hence in this
 approximation  equations (\ref{eq:tt}) (\ref{eq:rr}) and (\ref{eq:thetatheta})
take the following form, 
\begin{widetext}
\bea \label{eq:ttc}
& & 8 \pi \tilde{\mathcal{R}}(\gamma) [\frac{e^{-\lambda}}{r^2} 
( (\epsilon+p)\partial_r +
 \epsilon') W + (\epsilon+p) \frac{l(l+1)}{r^2} V ] P_l (\cos \theta)= 
 -\delta_s \epsilon(r) e^{i \omega t} P_{\tilde{l}}(\cos \tilde{\theta})
\eea
\bea \label{eq:rrc}
  8 \pi \tilde{\mathcal{R}}(\gamma)\{ [ \Gamma_1 p \frac{e^{-\lambda}}{r^2}
 \partial_r + \frac{p'}{r^2} e^{-\lambda}] W - \Gamma_1 p
 \frac{l(l+1)}{r^2} V ]\} P_l (\cos \theta)  = -\delta_s p(r) e^{i \omega t} 
P_{\tilde{l}} (\cos \theta).
\eea
\bea \label{eq:thetathetac}
& & 8 \pi \tilde{\mathcal{R}}(\gamma)\{ [ \Gamma_1 p \frac{e^{- \lambda}}{r^2}
 \partial_r + \frac{p'}{r^2} e^{-\lambda} ] W +  \Gamma_1 p
 \frac{l(l+1)}{r^2} V]\} P_l (\cos \theta) =  \delta_s p_{\tilde{l}}(r) 
e^{i \omega t} P_{\tilde{l}}(\cos \theta)
\eea
The solutions can easily be obtained for $W$ and $V$  using any two equations
from the above. Using equation (\ref{eq:ttc}) and (\ref{eq:rrc}) we get,  
\bea
W(r,t) P_l \cos(\theta) & =& - \frac{e^{\int \frac{1}{2} e^\lambda 
\{ \frac{\epsilon'}{(\epsilon+p)} + \frac{p'}{\Gamma_1 p} \}dr}}{\tilde{
\mathcal{R}}(\gamma)} \int e^{- \frac{1}{2} \int e^\lambda \{ 
\frac{\epsilon'}{(\epsilon+p)}+ \frac{p'}{\Gamma_1 p} \} dr''} 
 \frac{r^2}{ 8 \pi  (\epsilon + p)} 
\{\delta_s \epsilon(r) + \frac{(\epsilon+p)}{\Gamma_1 p)} \delta_s p(r) \} 
e^{i \omega t}  P_{\tilde{l}} (\cos \theta) dr'
\eea
\end{widetext}
Similarly an expression for $V$ can be obtained, while as we have mentioned
 earlier,
these perturbations are generalized stochastic in nature, hence they are
 meaningful only as ensemble averages. Thus $\langle W \rangle = \langle V 
\rangle= 0 $ and from the above expression we can easily evaluate the rms
 value and two point correlations in terms of $\delta_s p(r),\delta_s 
\epsilon(r) $. We see the appearance of $\tilde{\mathcal{R}}(\gamma)$ in the 
expression for stochastic $W$ and $V$, which represents susceptibility 
of the configuration to get perturbed, giving way to characterize the
dense matter properties in a statistical way. We have presented new results
here in a near-equilibrium configuration of a relativistic star in a 
perturbative way  through these foundational set of equations. 
We can see that these are not evolution equations for perturbations
obtained from the Euler equation, however they give the stochastic averages
and correlations in a statistical domain of study.  
\section{Conclusion and further directions}
In this article, we have presented a formalism to probe sub-hydro scale 
mesoscopic spherical polar perturbations  in a relativistic star, which
are of generalized stochastic nature.
We consider a  perfect fluid matter, which describes matter in  the 
 near-equilibrium  stages of collapse for a relativistic star.
The expressions worked out in this article are through a first principles
 approach which are important  to establish foundations for mesoscopic scale 
physics in order to understand the nature of
  dense matter in strong gravity regions. Further scope lies
in exploring the modes of  these stochastic perturbations, which will 
necessarily need numerical techniques to solve for  equations in
explicit form in order to determine them. This is our future endeavour, 
where for more realistic cases, we will  consider slowly rotating and
fast rotating  geometries eventually. The additional features that
these geometries would carry shall match with current interests in 
simulations and observations for neutron stars,  though at
 different
scales than are currently being explored. Recent efforts in sub-grid modelling
in numerical simulations for the interiors of compact dense matter stars
have started taking shape \cite{nils3}. This is a new opening to theoretical
 and mathematical modelling.  Hence the work presented in this article is a
 step closer towards this direction as 
a new analytical approach, though in a different context. We are 
thus moving towards more refined structures in the interiors of compact
objects. Since we are interested in deciphering the
behaviour of the dense fluid and the fluid degrees of freedom coupled to
metric perturbations, this gives the reason why we have explored the  polar
 perturbations in this article and ignored the axial perturbations.   
\section*{Acknowledgements}
The author is thankful to Rajesh Nayak for helpful discussions. This work
was funded by grant number DST/WoS-A/PM-03/2021 from DST India.

\section*{Appendix}
\begin{center}
\textbf{Review of polar perturbations in spherically symmetric star}
\end{center}

The polar perturbations  are even-parity perturbations on the 2-sphere of the
metric . The perturbed metric in Regge Wheeler gauge then is given by
\[ 
h_{\alpha \beta} = 
\begin{bmatrix} e^{2 \nu} H_0 Y_{lm} && H_1 Y_{lm} && 0 && 0 \\
Sym & & e^{2 \lambda } H_2 Y_{lm} &&  0 && 0 \\
Sym && Sym && r^2 K Y_{lm} && 0 \\
Sym & & Sym && Sym && r^2 K \sin^2 \theta Y_{lm} 
\end{bmatrix}
\]
Eventually the independent metric variables are commonly taken to be
$H_0, H_1,$and $K$ as it turns out that $H_2 =H_0$ using the Einstein's 
tensors $ \delta G^\theta_\theta - \delta G^\phi_\phi =0 $.   
The perturbed fluid four-velocity with $Y_{lm}$ replaced by the condition
$m=0$. 
\bea
\delta u^t &=& - e^{\nu} [ 1- \frac{1}{2} H_0 P_l(\cos \theta)] \\
\delta u^r & = & -\frac{e^{-(\nu+\lambda}}{r^2} W_{,t} P_l (\cos \theta) \\
\delta u^\theta  & = & e^{\nu}\frac{1}{r^2} V_{,t} \partial_\theta 
P_l(\cos \theta)\\
\delta u^\phi & = & 0
\eea
The Lagrangian change in  number density of baryons  is $\Delta n $, then
\be
\Delta n/n = \{-\frac{e^{-\lambda}}{r^2} W' - \frac{l(l+1)}{r^2}V + \frac{1}{2}
H_2 + K \} P_l (\cos \theta).
\ee 
The corresponding Eulerian changes in energy density and pressure are
\bea
\delta \epsilon & = & (\epsilon+p) (\Delta n/n) - \epsilon' e^{-\lambda} W 
P_l (\cos \theta) \\
\delta p & = & \Gamma p (\Delta n/n) - p' \frac{e^{-\lambda}}{r^2} W 
P_l(\cos \theta) .
\eea
 The Eulerian changes in the stress-energy tensor are
\bea
& & \delta T^t_t = - \delta \epsilon, \delta T^r_r = \delta T^\theta_\theta
= \delta T^\phi_\phi = \delta p , \\
& &  \delta T^r_t  = (\epsilon+p) u_t \delta u^r, \delta T^t_r = (\epsilon+ p)
 u_r \delta u^t, \\ 
& & \delta T^\theta_t = (\epsilon +p) u_t \delta u^\theta , 
\delta T^t_\theta = (\epsilon+ p) u_\theta \delta u^t
\eea
Rest of the components of $\delta T^\alpha_\beta $ vanish.  
\end{document}